\documentstyle[prl,twocolumn,aps]{revtex}

\newcommand{\extraspace}{\addtolength{\abovedisplayskip}{2mm} 
                        \addtolength{\belowdisplayskip}{2mm} 
                        \addtolength{\abovedisplayshortskip}{2mm} 
                        \addtolength{\belowdisplayshortskip}{2mm}} 
\newcommand{\be}{\begin{equation}\extraspace} 
\newcommand{\ee}{\end{equation}} 
\newcommand{\bea}{\begin{eqnarray}\extraspace} 
\newcommand{\eea}{\end{eqnarray}} 

\newcommand{\wtPsi}{\widetilde{\Psi}}
\newcommand{\del}{\partial}

\newcommand{\nonu}{\nonumber \\[2mm]} 
  
\begin{document}
\draft
\widetext
\title{New Class of Non-Abelian Spin-Singlet Quantum Hall States}
\author{Eddy Ardonne and Kareljan Schoutens}
\address{
Institute for Theoretical Physics,
Valckenierstraat 65, 1018 XE Amsterdam, THE NETHERLANDS \\
}
\date{November 25, 1998}
\maketitle
\begin{abstract}
We present a new class of non-abelian spin-singlet quantum Hall states,
generalizing Halperin's abelian spin-singlet states and the 
Read-Rezayi non-abelian quantum Hall states for spin-polarized electrons.
We label the states by $(k,M)$ with $M$ odd (even) for fermionic 
(bosonic) 
states, and find a filling fraction $\nu=2k/(2kM+3)$. The states
with $M=0$ are bosonic spin-singlet states characterized by an $SU(3)_k$ 
symmetry. We explain how an effective Landau-Ginzburg theory
for the $SU(3)_2$ state can be constructed. In general, the 
quasi-particles 
over these new quantum Hall states carry spin, fractional charge and 
non-abelian quantum statistics.

\end{abstract}

\pacs{PACS: 73.40.Hm, 71.10.-w, 73.20.Dx}

\vskip -4mm

%

%
%
%


The developments that followed the discovery of the fractional
quantum Hall effect in 1982 have challenged two traditional 
wisdoms on the nature of the quantum Hall states that are relevant for 
experimental observations. 
These two wisdoms concern the {\em spin}\ of the electrons that 
participate in a quantum Hall state, and the {\em quantum statistics}\ 
of the quasi-particle excitations over these states.

Starting with the first wisdom, it is evident that under the conditions
of the (fractional) quantum Hall effect, which happens in a strong
magnetic field,  there is an important Zeeman splitting between 
the energies of spin-up and spin-down polarized electrons.
One may therefore expect  that the observable quantum Hall states 
will be in terms of spin-polarized electrons. However, already in 1983
Halperin \cite{halp} pointed out that the energy associated with the 
Zeeman 
splitting is rather modest  as compared to other energy 
scales in the system. Because of this, quantum Hall states which are 
not spin-polarized but instead involve equal numbers of spin-up and 
spin-down electrons (forming a {\it spin-singlet}) are feasible.
The experimental confirmation of this idea came in 1989, when several
groups reported that the ground states at $\nu=4/3$ and $\nu=8/5$
are spin-unpolarized \cite{eis}. In recent experiments, which employed
hydrostatic pressure to reduce the $g$ factor, more detailed results
on quantum Hall spin-transitions were obtained \cite{wkang}.
In his 1983 paper \cite{halp}, Halperin proposed the following 
spin-singlet (SS) quantum Hall states 
\bea
\lefteqn{ \wtPsi_{\rm  SS}^{m+1,m+1,m}
(z_1,\ldots,z_N;w_1,\ldots,w_N) =}
\nonu && \quad
\Pi_{i<j}(z_i-z_j)^{m+1} \Pi_{i<j}(w_i-w_j)^{m+1} 
\Pi_{i,j}(z_i-w_j)^{m} 
\nonu &&
\label{eq:ss}
\eea
where $z_i$ and $w_i$ are the coordinates of the spin up
and spin down electrons, respectively. The state Eq.\
(\ref{eq:ss}) has filling fraction $\nu=2/(2m+1)$. 
Here and below we display
reduced quantum Hall wave functions $\wtPsi(x)$, which are related
to the actual wave functions $\Psi(x)$ via $\Psi(x)
= \wtPsi(x) \exp{(-\sum_i {|x_i|^2 \over 4 l^2}})$ with 
$x_i=z_i,w_i$ and $l=\sqrt{\hbar c \over e B}$ the magnetic 
length. It was emphasized in \cite{mr,bf} that the wave function
Eq.\ (\ref{eq:ss}) can be factorized into a charge factor times
a spin factor. The spin factor has an $SU(2)_1$ affine Kac-Moody
symmetry and describes semionic spinons that are also encountered
in other models of spin-charge separated electrons in $d=1+1$ 
dimensions. More general (abelian) spin-singlet states have
been described in the literature \cite{abelian-ss}.

Concerning the second wisdom, we remark that 
the traditional hierarchical quantum Hall states (Jain series) all 
share the property that the quantum statistics of their fundamental
excitations are fractional but abelian. While these states suffice to
explain the overwhelming majority of experimental observations,
there is the exception of the well-established quantum Hall state
at $\nu=5/2$, which does not fit into the hierarchical scheme. This
observation has prompted the analysis of new quantum Hall states, 
the Haldane-Rezayi state \cite{hr} and the $q=2$ pfaffian 
state \cite{mr} (both at $\nu=1/2$) being the most prominent among 
them. The quasi-hole excitations over the pfaffian quantum Hall 
states satisfy what is called {\it non-abelian statistics}
\cite{mr,nw}. By 
this one means that the wave function describing a number of 
quasi-holes at fixed positions has more than one component, and 
that the braiding of two quasi-holes is represented by a matrix 
that acts on this multi-component wave function. Since matrices
in general do not commute, such statistics are called non-abelian.
The non-abelian braid statistics of the quasi-holes over the
pfaffian quantum Hall states are reflected in their unusual
exclusion statistics, i.e., in the appropriate generalization of 
the Pauli Principle for particles of this type \cite{sc}.

In a recent paper \cite{rr}, Read and Rezayi proposed and studied a class
of spin-polarized, non-abelian quantum Hall states, that generalize
the pfaffian. Some of these states were independently considered
by Wen \cite{wen}. The most general non-abelian (NA) quantum
Hall 
state studied 
by Read and Rezayi, with labels $(k,M)$, is of the form
\bea
\lefteqn{\wtPsi_{\rm NA}^{k,M}(z_1,\ldots,z_N)=}
\nonu && \quad
\langle \psi(z_1) \ldots \psi(z_N) \rangle \,
\Pi_{i<j}(z_i-z_j)^{{2 \over k}+M}
\label{eq:rr}
\eea
with $\psi(z)$ a so-called order-$k$ parafermion and with the
brackets $\langle \ldots \rangle$ denoting a correlator in the 
associated conformal field theory (CFT). The physical picture
behind these wave functions is that of an instability involving 
a clustering of at the most $k$ particles, generalizing the 
notion of `pairing' that underlies the pfaffian states \cite{gww}.

In this Letter we describe a new class of quantum Hall states, 
which combine the feature of being a spin-singlet with that
of non-abelian statistics. These new states can be 
viewed as non-abelian generalizations of the spin-singlet states  
Eq.\ (\ref{eq:ss}), or,  alternatively, as spin-singlet analogues 
of the non-abelian states Eq.\ (\ref{eq:rr}). The wave function of
our most general non-abelian spin-singlet (NASS) state, labeled
as $(k,M)$ with $k>1$, is displayed in Eq.\ (\ref{eq:nass})
below. It has filling fraction
\be
\nu(k,M) = {2k \over 2kM+3} \ .
\label{eq:nu}
\ee
The simplest fermionic NASS state (with $k=2$, $M=1$) occurs at 
filling fraction $\nu=4/7$. In general,
the NASS states are competing with abelian spin-singlet states
that are possible at the same filling fractions Eq.\ (\ref{eq:nu}). 

In recent studies of the pfaffian and Read-Rezayi 
non-abelian quantum Hall states \cite{fntw,fns}, 
it has been emphasized that the 
essential mechanism of their non-abelian statistics is closely related 
to the presence of (a deformation of) a non-abelian $SU(2)_k$ 
affine Kac-Moody symmetry with $k>1$. In this Letter we shall see 
that, for the case of spin-singlet
non-abelian quantum Hall states, there is a very similar role for
a symmetry $SU(3)_k$ with $k>1$.

We start our presentation by a discussion of the non-abelian $SU(3)_1$
symmetry of a particular abelian spin-singlet quantum Hall state. 
In a recent paper \cite{fns}, it was emphasized that the bosonic 
Laughlin state at $\nu=1/2$ possesses
a non-abelian $SU(2)_1$ symmetry, which can be viewed as a continuous 
extension of the particle-hole symmetry at half-filling. We shall refer
to this symmetry as $SU(2)$-charge.  In earlier work by Balatsky and
Fradkin \cite{bf}, it was stressed that the  $(1/2,1/2,-1/2)$ Halperin
state, which is a spin-singlet state at $\nu=\infty$, i.e., at ${\bf B}=0$,
possesses $SU(2)_1$ non-abelian symmetry, which we here call
$SU(2)$-spin. 
Combining these two observations, one expects to
find a bosonic spin-singlet quantum Hall state at {\em finite}\  
$\nu$, in which $SU(2)$-charge and $SU(2)$-spin combine into a 
non-trivial extended symmetry. The nature of this extended 
symmetry can be traced by analyzing the algebraic properties of
creation and annihilation operators for spin-full hard-core bosons. 
We define
\begin{eqnarray}
B^\dagger_{\sigma} = \psi_{1\sigma}^\dagger \psi_2 \ ,
&& \qquad
B_{\sigma} = \psi_2^\dagger \psi_{1\sigma} 
\nonu
B^3_a=\psi_{1\sigma}^\dagger \sigma_{\sigma \rho}^a \psi_{1\rho} \ ,
&& \qquad
B^3=\psi_{1\sigma}^\dagger \psi_{1\sigma} - 2\psi_2^\dagger \psi_2 
\label{eq:Bs}
\end{eqnarray}
with $\sigma,\rho=\uparrow,\downarrow$ and
with $\psi^\dagger_{1\sigma}$, $\psi_{1\sigma}$ and $\psi_2^\dagger$,
$\psi_2$ the creation
and annihilation operators of spin-$1/2$ bosonic particles and holes. 
[We remark that in this set-up the holes do {\it not}\
carry a spin index.] The `hard-core' condition is implemented by the 
constraint 
\begin{equation}
\psi_{1\sigma}^\dagger \psi_{1\sigma} +\psi_2^\dagger \psi_2=1 \ .
\end{equation}
Using the defining commutators 
\be
\left[ \psi_{1\sigma} , \psi_{1\rho}^\dagger \right] =
\delta_{\sigma \rho} \ , \qquad 
\left[ \psi_2 , \psi_2^\dagger \right] = 1 
\label{eq:comms}
\ee
one shows that the eight operators $B^A$ of Eq.\ (\ref{eq:Bs})
form an adjoint representation of the algebra $SU(3)$. In
standard mathematical notation, we denote by
$B_\alpha$ the element of the Lie algebra $SU(3)$ that
corresponds to a root $\alpha$. With simple roots 
$\alpha_1= (\sqrt{2},0)$, $\alpha_2=(-\sqrt{2}/2,\sqrt{6}/2)$ 
we identify the boson creation operators as
\be
B^\dagger_{\uparrow}=B_{\alpha_1} \ , \quad
B^\dagger_{\downarrow}=B_{-\alpha_2} \ .
\label{eq:Balpha}
\ee
We conclude that the kinematics of hard-core spin-full
bosons are organized by an $SU(3)$ symmetry. In its
fermionic incarnation $SU(2|1)$, this symmetry is well
known from the supersymmetric $t$-$J$ model \cite{tJref}.

Alerted by this result, we quickly find that the
$(2,2,1)$ bosonic spin-singlet state (Eq.\ (\ref{eq:ss})
with $m=1$) possesses an $SU(3)_1$ 
global symmetry. One way to see this is by recognizing that
the inverse $K$-matrix, given by
\begin{equation}
K^{-1} =
 {1 \over 3} \left(\begin{array}{cc}
      2 & -1 \\ -1 & 2 \end{array} \right) \ ,
\end{equation}
is equal to the inverse Cartan
matrix of $SU(3)$, up to a trivial change of sign. 
Working out the $SU(3)$ structure of the edge 
CFT for the $(2,2,1)$ state, one identifies the Cartan
subalgebra generators $B^3$ and $B^3_3$ with the 
spin and charge bosons according to
\be
B^3 = i \, \sqrt{6} \, \del\varphi_c \ , \qquad
B^3_3 = \, -i \sqrt{2} \, \del\varphi_s \ .
\ee
The fundamental quasi-particles reside in the triplet ${\bf 3}$ 
and anti-triplet ${\bf \bar{3}}$ representations of $SU(3)$, 
with spin and charge quantum numbers
\bea
\phi^1 \ : & \quad {\rm spin} \ \uparrow & , \quad q=-1/3
\nonu
\phi^2 \ : & \quad {\rm spin} \ \downarrow & , \quad q=-1/3
\nonu
\phi^3 \ : & \quad {\rm spin} \ \, 0 & , \quad q=2/3
\label{eq:tripl}
\eea
and opposite for the anti-triplet.
Following \cite{es}, we can construct the complete edge theory 
in terms of multi-particle states consisting 
of quanta of the fields
$\phi^i$, $i=1,2,3$.  The systematics of this construction lead to
a notion of `fractional exclusion statistics' of these quasi-particles. 
In \cite{bs}, the mathematical details of these fractional statistics,
which differ from those proposed by Haldane \cite{hald}, were 
presented.
As a direct application of the results of \cite{bs}, we may recover 
the Hall conductance $\sigma_H$ of the $(2,2,1)$ state by working out 
the following expression \cite{es}
\be 
\sigma_H = n^{\rm max} q^{2} \, {e^2 \over h} \ ,
\ee
with $n^{\rm max}$ equal to the maximal occupation of
a given single quasi-particle state.
Substituting the values $q=2/3$, $n^{\rm max}=3/2$ for the
positive-charge carriers $\phi^3$, we recover the value 
$\sigma_H={2 \over 3} {e^2 \over h}$.

An effective Landau-Ginzburg bulk theory for the $(2,2,1)$ 
state can be cast in the following form
\be
{\cal L} = |D B|^2 + V(B) + {\cal L}_{\rm CS}(a) 
+ \epsilon^{\mu \nu \lambda} A_\mu^{\rm ext} f_{\nu \lambda}^3
\label{eq:lg1}
\ee
where $B$ is the $SU(3)$ octet Bose field, $D B$ the covariant
derivative in the adjoint representation, $V(B)$ a potential, 
and $a^A_\mu$ an $SU(3)$ Chern-Simons (CS) gauge field. 
The external field $A^{\rm ext}_\mu$ couples to the $f^3$ 
component of the field tensor of the gauge field $a^A_\mu$.
The Chern-Simons lagrangian is given by
\begin{equation}
{\cal L}_{\rm CS}(a) = 
\frac{1}{4\pi} {\epsilon^{\mu\nu\lambda}}\
\left({a_{\mu}^A}{\partial_\nu}{a_{\lambda}^A} + \frac{2}{3}
{f_{ABC}}{a_{\mu}^A}{a_{\nu}^B}{a_{\lambda}^C}\right)\ .
\end{equation} 
For the justification of the result Eq. (\ref{eq:lg1}) we refer to 
\cite{fns}, where an analogous result for the $SU(2)_1$ 
invariant $\nu=1/2$ state was presented.

Having understood the $SU(3)_1$ structure of a particular abelian
SS state, we can proceed with the construction of NASS states.
Closely following the logic presented in \cite{fns}
(see also \cite{cappelli}),  we first consider
a state with symmetry $SU(3)_2$, which we obtain by performing
a (dual) coset reduction on two copies of  the $SU(3)_1$ state.
As explained in \cite{fns}, this reduction procedure 
leads to a statistical transmutation of one of the octet fields and
renders the statistics of the triplet quasi-particles $\phi^i$
non-abelian.

The effective edge CFT for the $SU(3)_2$ theory is completely
determined by the known structure of the $SU(3)_2$ chiral 
Wess-Zumino-Witten (WZW) theory. At the same time, this 
CFT can be used to generate an explicit expression for the ground 
state wave function in the bulk (see Eq.\ (\ref{eq:nass}) below), 
and for the 
wave functions representing various quasi-particle excitations. 
The spin and charge quantum numbers of the fundamental triplet 
excitations are the same as those listed in Eq.\ (\ref{eq:tripl}), 
but the exclusion statistics are now non-abelian. 

Before presenting more general NASS states, we remark that an
effective Landau-Ginzburg theory for the $SU(3)_2$ NASS state
is readily given, by generalizing the construction of \cite{fns}
for the bosonic pfaffian state with symmetry $SU(2)_2$.
The Landau-Ginzburg theory  is obtained by taking two copies of 
the $SU(3)_1$ theory Eq.\ (\ref{eq:lg1})
and using a pairing mechanism that is similar
to the electron pairing in a Bardeen-Cooper-Schrieffer 
theory. The pairing induces
a symmetry breaking  $SU(3)_1 \times SU(3)_1 \to SU(3)_2$
and hence induces a level $k>1$ non-abelian symmetry that
is characteristic of non-abelian statistics. 
One may check that the stable 
vortices in the broken-symmetry theory correspond to the 
quasi-particles that are identified using the edge CFT.


We now turn to the description of a two-parameter family of 
NASS states. They are obtained by taking a back-bone $SU(3)_k$
theory, dressed with an additional Laughlin factor with
exponent $M$. To obtain explicit expressions for the
corresponding wave functions, we rely on the $SU(3)$ 
parafermions introduced by Gepner in \cite{gep}. These
parafermions, written as $\psi_{\alpha}$, are labeled by 
{\em roots}\ $\alpha$ of $SU(3)$ and have the property that
$\psi_\alpha=\psi_\beta$ when $\alpha-\beta$ is an element
of $k$ times the root lattice. In terms of the $\psi_\alpha$
and of two auxiliary bosons $\varphi_{1,2}$, the affine
currents $B_\alpha(z)$ at level $k$ are written as
\be
B_\alpha(z) \propto \psi_{\alpha} 
            \exp(i \alpha \cdot \varphi / \sqrt{k})(z) \ .
\label{eq:Bpara}
\ee
Using the identification Eq.\ (\ref{eq:Balpha}), we 
arrive at the following expression for the NASS state
associated to $SU(3)_k$
\bea
&& \lim_{z_\infty \to \infty} (z_\infty)^{\frac{6N^2}{k}} \, 
\langle
\exp(i\frac{N}{\sqrt{k}}(\alpha_2-\alpha_1)\cdot\varphi)(z_\infty) 
\nonu && \ \times \,
B_{\alpha_1}(z_1)\ldots B_{\alpha_1}(z_N)
B_{-\alpha_2}(w_1)\ldots B_{-\alpha_2}(w_N)
\rangle \ .
\eea
Substituting the form Eq.\ (\ref{eq:Bpara}), one 
observes that the correlator factorizes as a parafermion
correlator times a factor coming from the vertex
operators. The later is seen to combine into the
$1/k$-th power of the $(2,2,1)$ spin-singlet
wave function. Multiplying the result with an overall 
Laughlin factor, we arrive at the 
following wave function for the $(k,M)$ NASS state
\bea
&& \wtPsi_{\rm NASS}^{k,M}(z_1,\ldots,z_N;
   w_1,\ldots,w_N)=
\nonu && \quad
\langle 
\psi_{\alpha_1}(z_1) \ldots \psi_{\alpha_1}(z_N) 
\psi_{-\alpha_2}(w_1) \ldots \psi_{-\alpha_2}(w_N) 
\rangle 
\nonu && \quad \times
\left[ \wtPsi_{\rm SS}^{2,2,1}(z_i;w_j) \right]^{1/k} \,
\wtPsi^M_{\rm L}(z_i;w_j) \ ,
\label{eq:nass}
\eea
with $\wtPsi^{2,2,1}_{\rm SS}$ as in Eq.\ (\ref{eq:ss}) and
$\wtPsi_{\rm L}^M$ the standard Laughlin wave function
with exponent $M$, with odd (even) $M$ giving
a fermionic (bosonic) state. By combining the final two 
factors in Eq.\ (\ref{eq:nass}) one recognizes a two-layer 
state with 
label $(M+2/k,M+2/k,M+1/k)$, and since it is these 
factors that determine the filling fraction $\nu$
we immediately derive the value given in 
Eq.\ (\ref{eq:nu}).

For $k=1$, the state Eq.\ (\ref{eq:nass}) reduces to
the abelian SS state Eq.\ (\ref{eq:ss}) with $m=M+1$.
For $k>1$, the parafermion correlator is non-vanishing
for $N$ an integer multiple of $k$. The simplest 
non-trivial example 
of a wave function of the type Eq.\ (\ref{eq:nass}) 
is the case $(k=2,M=0)$ for a total of $N=4$ 
bosonic particles. We find
\bea
&& \wtPsi^{k=2,M=0}_{\rm NASS}(z_1,z_2;w_1,w_2) =  
\nonu
&& \quad 2( z_1 z_2 + w_1 w_2) - (z_1+z_2)(w_1+w_2) \ .
\label{eq:4part}
\eea
By inspecting the zeroes of this wave function, we can
understand the pairing that underlies this particular
quantum Hall state (compare with \cite{mr,gww,rr}).
We see that upon sending $z_2\to z_1$ or 
$w_1 \to z_1$ the wave function Eq.\ (\ref{eq:4part}) 
does {\it not}\ go to zero. However, as soon as three
or more particles come together, we do get a zero.
(For three particles of the same spin, i.e., three
$z_i$ or three $w_i$, this can not be seen from 
Eq.\ (\ref{eq:4part}); it follows however from the 
Operator Product Expansion
structure of the $SU(3)$ parafermions 
$\psi_{\alpha}$.) We conclude that the pairing of the 
$k=2$ NASS states is similar to that of the pfaffian
states for spin-polarized electrons. By the 
analogy with the findings of \cite{rr}, one similarly
expects that the instability underlying the level-$k$ 
NASS states for $k>2$ will be a `$k$-particle clustering'. 

The CFT underlying the states Eq.\ (\ref{eq:nass})
is unitary, and the bulk-edge correspondence for these
states thus avoids some of the subtleties that arise
for the Haldane-Rezayi NASS state. It is therefore 
straightforward
to derive the spin and charge quantum numbers of the
fundamental quasi-particles over these states, and 
to determine exponents for quasi-particle and electron
edge tunneling processes. For $M\neq 0$ the $SU(3)_k$
symmetry is broken and $SU(3)$ quantum numbers are no 
longer meaningful. The fundamental flux-${1 \over 2k}$
quasi-holes carry charge $q=\pm 1/(2kM+3)$ and spin
$1/2$. Their conformal dimension, which is obtained
by adding contributions from the parafermion sector
and from the spin and charge sectors, equals
\be
\Delta_{\rm qh} = 
{(5k-1)M+8 \over 2(k+3)(2kM+3)} \ .
\label{delta}
\ee 
The edge electrons have charge $-1$, spin $1/2$ and 
conformal dimension $\Delta_{\rm el}=(M+2)/2$,
independent of $k$. The non-abelian braid and
exclusion statistics of the various quasi-particles 
follow by a straightforward generalization of the
techniques of \cite{mr,nw,sc,bs}.

We remark that some of the filling factors 
for which fermionic NASS states exist agree with
values for which spin-transitions have been seen in 
experiments. 
With regard to the competition between abelian and 
non-abelian spin-singlet quantum Hall states with
the same filling fraction $\nu$, we remark the 
following.
Based on explicit numerical work, the authors of
\cite{rr} have suggested that in the second Landau
level non-abelian spin-polarized states 
tend to be favored over their abelian counterparts.
By analogy, one may expect that the NASS states 
proposed in this Letter will be favored over
abelian SS states when $\nu>2$.
On the basis of the reasoning presented in \cite{gww}
one also expects that the $k=2$ NASS states will be 
particularly relevant for samples with wide well or 
double well geometries.
We finally remark that, experimentally, one can in
principle distinguish between abelian and non-abelian
quantum Hall states by studying processes where a
current tunnels through the quantum Hall medium. Such
experiments probe conformal dimensions such as Eq.
(\ref{delta}), and these in general 
differ between abelian and non-abelian states with the 
same filling fraction.

\vskip 5mm

\noindent
{\it Acknowledgements.}\ 
We thank Bernd Schroers for illuminating discussions. KS
thanks Nick Read and Ed Rezayi for discussions on 
non-abelian quantum Hall states and Eduardo Fradkin and
Chetan Nayak for collaboration on a closely related 
project. This research is supported in part by the 
Foundation FOM of the Netherlands.

\end{document}